\documentclass[conference]{IEEEtran}
\usepackage{ amssymb }
\usepackage{color}
\usepackage{textcomp}
\usepackage{algorithm}
\usepackage{algpseudocode} 
\usepackage{tikz} 
\usepackage{cite}
\usepackage{mathtools}

\DeclareMathOperator*{\argmin}{arg\,min}

\usepackage{amsthm}
\usepackage{amsmath}
\newtheorem{theorem}{Theorem}

\newtheorem{proposition}[theorem]{Proposition}

\newtheorem{definition}{Definition}

\title{GDSP: A Graphical Perspective on the Distributed Storage Systems}

\author{\IEEEauthorblockN{Saeid Sahraei\IEEEauthorrefmark{1},
Michael Gastpar\IEEEauthorrefmark{2}}
\IEEEauthorblockA{School of Computer and Communication Sciences,
EPFL\\
Lausanne, Switzerland\\
Email: \IEEEauthorrefmark{1}saeid.sahraei@epfl.ch,
\IEEEauthorrefmark{2}michael.gastpar@epfl.ch}}

\begin{document}

\maketitle 
\begin{abstract}
The classical distributed storage problem can be modeled by a k-uniform {\it complete} hyper-graph where vertices represent servers and hyper-edges represent users. Hence each hyper-edge should be able to recover the full file using only the memories of the vertices associated with it. This paper considers the generalization of this problem to {\it arbitrary}  hyper-graphs and to the case of multiple files, where each user is only interested in one, a problem we will refer to as the graphical distributed storage problem (GDSP). Specifically, we make progress in the analysis of minimum-storage codes for two main subproblems of the GDSP which extend the classical model in two independent directions: the case of an arbitrary graph with multiple files, and the case of an arbitrary hyper-graph with a single file.
\end{abstract}

\section{Introduction}

The Distributed Storage Problem (DSP) has received a lot of attention in the recent literature,
with a key focus on the important trade-off between storage and repair bandwidth 
\cite{dimakis2010network,dimakis2011survey,rashmi2011optimal,weatherspoon2002erasure,rhea2001maintenance,rodrigues2005high,papailiopoulos2014locally,rhea2003pond}.
In this paper, we address a novel extension of the DSP.
To put this in perspective, note that the classical DSP can be viewed as a complete $k$-uniform hyper-graph with $K$ vertices which represent the servers and ${K\choose k}$ hyper-edges representing the users which are connected to the corresponding set of servers. Each vertex of the graph is equipped with a memory of size $M$ and each hyper-edge should be able to recover a specific file $A$ from the memories of the vertices associated with him. A minimum-storage code (which is the focus of the current work) minimizes the total required memory $MK$ under these constraints. 

More generally, suppose we have an arbitrary hyper-graph defined over $K$ vertices and a set of $N$ independent files $\{A_1,\dots,A_N\}$. Each hyper-edge of the graph is colored by some $c\in\{1,\dots,N\}$. We have an entire memory budget of $M$ which we distribute among different vertices of the hyper-graph. To each vertex $i$ we assign a memory of size $M_i$ where $\sum_i M_i \le M$. Each server stores a function $h_i$ of the files in his memory such that $H(h_i)\le M_i$. These functions must be designed such that if $S=\{s_1,\dots,s_\ell\}$ is a hyper-edge of the graph colored with $c$, then $H(A_c|h_{s_1},\dots,h_{s_\ell}) = 0$. The question is what is the minimum total memory budget $M$ which allows us to accomplish this task for a particular colored hyper-graph. We will refer to this as the {Graphical Distributed Storage Problem (GDSP)}. We will establish close connections between this problem and well known problems in the networking literature. These connections also imply that the problem in its full generality is too difficult to tackle at once. Therefore, we focus in this work on two sub-models, extending the classical DSS model in two orthogonal directions. Firstly, and for most of the paper, we will study a graph (a hyper-graph where size of each hyper-edge is two) where edges are colored arbitrarily. In Section \ref{sec:simple}, we will propose an achievability strategy for a - practically motivated - subclass of such graphs which we refer to as ``smoothly colored graphs". We will show that under certain constraints our strategy is optimal. We consider this proof of optimality as the main contribution of the paper. 
Secondly, we will briefly investigate a hyper-graph in the presence of only one file. For such a hyper-graph in Section \ref{sec:hyper}  we will characterize the minimum total required memory as the solution to an LP.  

To motivate the extension of the distributed storage problem considered in this paper, let us start by reconsidering
the classical version. There, a key requirement is that {\it every}  sufficiently large subset of the servers must enable full recovery of the entire file.
For several scenarios of potential practical interest, this requirement could be unnecessarily stringent. 
Consider for example a setting with different classes of servers. Some servers could be more
powerful than others, or more reliable. Then, a natural consideration would be to suppose that every file recovery
would always involve at least one of the more powerful servers. In other words, one would not impose a requirement
that a subset consisting of only less powerful servers must enable file recovery. This naturally leads to a more general
hyper-graph model, beyond the uniform complete ones studied in the classical setup.
A practical framework where such a combination of more and less powerful servers might appear are caching networks
for content distribution. In such networks, there are auxiliary servers that help speed up data delivery.
However, it will generally not be possible to fully recover the desired content only from auxiliary servers. Rather, an additional
call to one of the (more powerful, but typically overloaded) main servers will be necessary.

The second generalization of our work concerns the file itself: In the classical problem, there is a single file.
In our extension, we allow for several files, and each user is requesting only one of the files. 
Again, such a scenario is of potential practical interest, for example, in a geographical setting:
let us suppose for the sake of argument that the servers are geographically distributed in a large area,
and let us envision content distribution that is location-specific, as in many of the commercial video distribution services.
Here, some servers will serve only one geographical sub-area while other servers will serve multiple,
leading to a hyper-graph where each hyper-edge will potentially seek to recover
a different file, specific to the geographical location.

\subsection{Connections with the existing literature}

\label{sec:nif}
\begin{figure}
\includegraphics[scale=0.5]{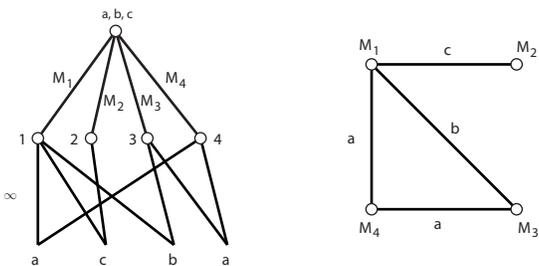}
\caption{What is the tradeoff between $M_1, M_2, M_3$ and $M_4$ such that the network admits a rate of $1$?}
\label{fig:flow}
\end{figure}

Our problem can be seen as a special instance of the single source {network information flow problem.} The source is connected to $K$ intermediate nodes with links with capacities $M_1,\dots,M_K$ respectively. These nodes represent the vertices in the GDSP. We have $|E|$ sinks corresponding to the hyper-edges of the GDSP. For any hyper-edge $S=\{s_1,\dots,s_\ell\}$, we connect  all the intermediate nodes $s_1,\dots,s_\ell$ to the corresponding sink with links with infinite capacities. Our problem is equivalent to finding the minimum sum rate $\sum_{i=1}^K M_i$ such that, given proper manipulation of the data at the intermediate nodes, each sink can recover its desired message at a rate of $1$ bit per second (see Figure \ref{fig:flow} for an illustration). For the network in Figure \ref{fig:flow} one can use Theorem \ref{thm:onesided} to show that the minimum sum achievable rate is $M^* = 3$.

\begin{figure}
\includegraphics[scale=0.6]{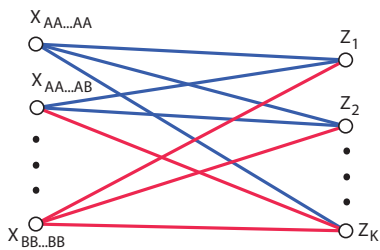}
\caption{The caching problem can be viewed as a complete bipartite graph. Blue edges must be able to recover file $A$ and the red ones, file $B$.}
\label{fig:MN}
\end{figure}

In \cite{maddah2014fundamental} the problem of centralized coded caching has been studied which can be briefly described as follows. There is a network consisting of $K$ users and one server. Each user is equipped with a memory of size $M$. The server has access to $N$ independent files and each user is interested in precisely one of these files. The goal is for the server to transmit these files to the users in two communication phases. A placement phase, where the server transmits a private message of size $M$ to each user {\it without } any prior knowledge of the interests of the users; and a delivery phase where the server, after learning the requests of the users, transmits one broadcast message of size $R$ to simultaneously satisfy all the requests. The challenge is to find a trade-off between these two rates, $R$ and $M$. This problem can be equivalently represented by a GDSP defined over a complete bipartite graph where on one side we have $K$ vertices standing for the user memories and on the other side we have $N^K$ vertices representing the delivery messages (see Figure  \ref{fig:MN}). While in general one can study the overall trade-off among $(M_1,\dots,M_K,R_1,\dots,R_{N^K})$, the authors in \cite{maddah2014fundamental} are limiting their analysis to $M_{i} = M$ and $R_j = R$ whereas we are interested in minimizing $\sum_i M_i + \sum_j R_j$.\\

There are also close connections between our work and FemtoCaching \cite{shanmugam2013femtocaching} where caching over an arbitrary hyper-graph is studied. The vertices and edges in GDSP correspond to femto-cells and users in the FemtoCaching model, respectively. There are however several differences. To mention a few, in the FemtoCaching model: all the caches are assumed to be of the same size; the file requests per hyper-edge are unknown and are modeled by a popularity distribution (which does not vary across different users) and last but not least coding across different files is not permitted and the main quest is to find the ``best" caching strategy under this restriction.  


\section{Graphs with Multiple Files}
\label{sec:simple}
Let us first formally define the problem. We have a set of $N$ independent files $\{A_1,\dots,A_N\}$ where $A_i$ consists of $F$ independent and uniformly distributed symbols over ${\mathbb F}_q$ where $q$ is a sufficiently large prime number. We have a colored graph $G = ({\cal V},E)$ where ${\cal V} = \{v_1,\dots,v_K\}$ is a set of $K$ vertices and $E$ is a set of tuples of the form $(\{i,j\},c)$ where $i,j\in \{1,\dots,K\}$, $ i\neq j$ and $c\in{\cal N} = \{1,\dots,N\}$. For any $\{i,j\}$ there is at most one such tuple in $E$, that is, if $(\{i,j\},c)\in E$ and $(\{i',j'\},c')\in E$ and $c\neq c'$ then $\{i,j\}\neq \{i',j'\}$. The parameter $c$ specifies the color of the edge or the file that the edge is interested in. Each vertex $i$ of the graph is equipped with a memory of size $M_{i,F}F$ where he stores a function the files, that is, $h_{i,F} = h_{i,F}(A_1,\dots,A_N)$ such that both conditions below are satisfied.
\begin{equation}
H(h_{i,F})\le M_{i,F}F
\label{eqn:condition1}
\end{equation}
\begin{equation}
H(A_c|h_{i,F} , h_{j,F}) = 0\;\; \mbox{ for all }(\{i,j\},c)\in E.
\label{eqn:condition2}
\end{equation}
Note that all the entropy terms are calculated base $q$. For a given colored graph ${G}$, we say that a memory allocation $(M_{1,F},\dots,M_{K,F})$ is valid if there exists functions $h_{i,F}(\cdot)$ that satisfy \eqref{eqn:condition1} and \eqref{eqn:condition2}. In this case, we call $(h_{1,F}(\cdot),\dots,h_{1,F}(\cdot))$ a valid assignment too. We say that a normalized sum rate of $M$ is achievable if there exists a sequence $\{(M_{1,F},\dots,M_{K,F})\}_{F=1}^\infty$ such that $(M_{1,F},\dots,M_{K,F})$ is a valid assignment for all $F$ and 
\begin{equation}
\lim_{F\rightarrow\infty}\sum_{i=1}^KM_{i,F} \le M.
\end{equation}
 Our goal is to find the minimum normalized sum achievable rate $M^*$ for a given colored graph ${G}$. That is $M^* = \inf\{M\Big| M \mbox{ is achievable}\}$. When clear from the context, we omit the subscript $F$ from $M_{i,F}$ and $ h_{i,F}$ to simplify the notation.

Motivated by the arguments in the introduction, let us now introduce a model which we refer to as a ``smoothly colored graph". Intuitively, a smoothly colored graph is one that can be ``partitioned" into several clusters each of which representing a certain geographical location.  The edges connecting the vertices within each cluster are colored differently from the other clusters, whereas the edges that connect vertices from two different clusters can be colored similarly to either of the two clusters. Such cross edges represent users which have access to servers from both clusters. Let us define this concept more formally.

\begin{definition}[Smoothly Colored Graphs] We say that a graph $G=({\cal V},E)$ is smoothly colored with respect to a partitioning ${\cal N}_1, {\cal N}_2,\dots,{\cal N}_L$ of ${\cal N}$ if the set ${\cal V}$ can be partitioned into $L$ subsets $ {\cal V}_1,{\cal V}_2,\dots,{\cal V}_L$ such that if $v_i,v_j\in {\cal V}_\ell$ and $(\{i,j\},c)\in E$ then $c \in {\cal N}_\ell$ and if $v_i\in {\cal V}_\ell$ and $v_j\in {\cal V}_{\ell'}$ and $(\{i,j\},c)\in E$ then $c\in {\cal N}_\ell\cup{\cal N}_\ell'$. For $i,j\in\{1,2,\dots,L\}$, we represent by ${\cal F}_{i,j}$ the subset of vertices in $ {\cal V}_j$ which are connected to at least one vertex outside of ${\cal V}_j$ with an edge colored with some $c\in {\cal N}_i$. Formally,

\begin{eqnarray*}
{\cal F}_{i,j} &=& \left\{v\in {\cal V}_j\Big| \exists u\notin {\cal V}_{j}, \;c\in{\cal N}_i \;,\;s.t.\;\; (\{u,v\},c)\in E\right\},\\ &&\forall i,j\in \{1,\dots,L\}.
\end{eqnarray*}
\end{definition}

\begin{figure}
\includegraphics[scale=0.35]{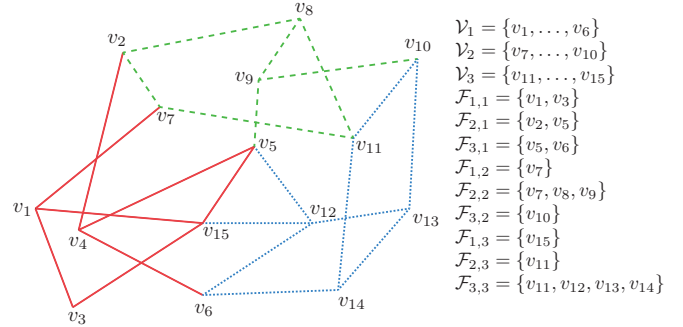}
\caption{A smoothly colored graph with three clusters.}
\label{fig:smooth}
\end{figure}

An example of a smoothly colored graph with three clusters and $|{\cal N}_1|=|{\cal N}_2|=|{\cal N}_3| =1$ has been depicted in Figure \ref{fig:smooth}. 

Our approach is to reduce the GDSP over a smoothly colored graph to several smaller instances of GDSP over subgraphs representing different clusters. This can be interesting for several reasons. Firstly, if each cluster is colored with only one color, we can use proposition \ref{prop:onlysizematters} from Section \ref{sec:hyper} in order to provide an exact solution for each cluster and consequently, find the exact solution for the overall network. Secondly, even within the realm of linear codes (for a fixed $F$), the complexity of an exhaustive algorithm grows exponentially with $N$. Therefore, any preprocessing that reduces $N$ can significantly improve the running time of the overall algorithm. 

Suppose $\mbox{solve}(G)$ is an optimal algorithm that given a graph $G$ returns any valid assignment $(M^*_1,\dots,M^*_K)$ for which $\sum_{i=1}^K{M^*_i} = M^*$. Consider now Algorithm \ref{Algorithm1} which given a graph $G$ and an arbitrary partitioning of the colors into ${\cal N}_1,\dots,{\cal N}_L$ returns $\mbox{SUP}(G,{\cal N}_1,\dots,{\cal N}_L)$, a superposition of $\mbox{solve}(G_1),\dots,\mbox{solve}(G_L)$ where $G_\ell$ is a subgraph of $G$ which only retains the edges colored by $c\in {\cal N}_\ell$ and eliminates all the other edges.

\begin{algorithm}
\caption[caption]{Superposition Algorithm}
\begin{algorithmic}[1]
\Statex {{\bf Input:} $G =({\cal V},E)$ and a partitioning of colors into ${\cal N}_1,\dots,{\cal N}_L$.}
 \Statex {\bf Output: }{$(M_{1},\dots,M_{K})=$ SUP$(G,{\cal N}_1,\dots,{\cal N}_L)$}
\State {Construct the subgraphs $G_\ell = ({\cal V},E_\ell)$ such that $(\{i,j\},c)\in E_\ell$ if and only if $(\{i,j\},c)\in E$ and $c\in {\cal N}_\ell$.}
\State {Let $(M^{(\ell)}_1,\dots,M^{(\ell)}_K) = solve(G_\ell)$ for all $\ell\in\{1,\dots,L\}$}.
\State {\Return $(\sum_{\ell=1}^L M^{(\ell)}_1,\dots,\sum_{\ell=1}^L M^{(\ell)}_K)$.}
\end{algorithmic}
\label{Algorithm1}
\end{algorithm}

The following theorem tells us that for a smoothly colored graph with $|{\cal N}_i| = 1$, and under the constraint ${\cal F}_{k,j}\cap {\cal F}_{\ell,j} = \varnothing$ for $k\neq \ell$, Algorithm \ref{Algorithm1} returns an exact solution.
\begin{theorem}
Suppose a graph $G = ({\cal V},E)$ is smoothly colored with respect to ${\cal N}_1,\dots,{\cal N}_L$ where $|{\cal N}_\ell|=1$ for all $\ell$. Suppose further that ${\cal F}_{k,j}\cap {\cal F}_{\ell,j} = \varnothing$ for all $j,k,\ell\in \{1,\dots,L\}$, $k\neq \ell$. Let $(\hat{M}_1,\dots,\hat{M}_K) = \mbox{SUP}(G,{\cal N}_1,\dots,{\cal N}_L)$ and $({M}^*_1,\dots,{M}^*_K) = \mbox{solve}(G)$ and, $\hat{M} = \sum_{i=1}^K \hat{M}_i$ and $M^* = \sum_{i=1}^K {M}^*_i$. We have $\hat{M} = M^*$.
\label{thm:onetoone}
\end{theorem}




\begin{IEEEproof}
Suppose $(h_1^*,\dots,h_K^*)$ is a valid assignment for $G$ which satisfies $H(h_\ell^*) \le M_\ell^*$.
Without loss of generality let us assume ${\cal N}_\ell = \{\ell\}$. Consider the following memory assignment for $G_\ell$. For all $v_i\in {\cal V}_\ell \backslash \bigcup_{k\neq \ell}{\cal F}_{k,\ell}$ we set ${M}^{(\ell)}_{i} = M^*_i$. For all $v_i\in  {\cal F}_{k,\ell}$ for $k\neq\ell$, we set $M^{(\ell)}_i = M_i^* - (1 - \min_{j} M^*_j)^+$ where the minimum is over all $j$ such that $v_j\in {\cal V}_k$ and $(\{i,j\},k)\in E$. (Note that $x^+$ stands for $\max\{x,0\}$). For all $v_i\in {\cal F}_{\ell,k}$ we set ${M}^{(\ell)}_{i} = (1 - \min_j M^*_j)^+$ where the minimum is over all $j$ such that $v_j\in {\cal V}_\ell$ and $(\{i,j\},\ell)\in E$. Finally, for all $v_i\in {\cal V}\backslash\left({\bigcup_{k\neq \ell}{\cal F}_{\ell,k} \cup {\cal V}_\ell}\right)$ we set  ${M}^{(\ell)}_i = 0$. We first prove that this is a valid assignment for $G_\ell$. Since $|{\cal N}_\ell| = 1$, by proposition \ref{prop:onlysizematters} we only need to prove that ${M}^{(\ell)}_{i} + {M}^{(\ell)}_{j}\ge 1$ if $(\{i,j\},\ell)\in E$. We consider several different cases.
\begin{itemize}
\item{$i,j\in {\cal V}_\ell \backslash \bigcup_{k\neq \ell}{\cal F}_{k,\ell}$. In this case we have ${M}^{(\ell)}_i  + {M}^{(\ell)}_j =  M^*_i + M^*_j $. Since $(M^*_1,\dots,M^*_K)$ is a valid assignment for $G$, we must have that $M^*_i + M^*_j \ge 1$, by proposition \ref{prop:onlysizematters}.}
 
 \item{$i\in {\cal V}_\ell\backslash \bigcup_{k\neq \ell}{\cal F}_{k,\ell}$ and $j\in {\cal F}_{\ell,k}$ for some $k\neq \ell$.}
 
 We have ${M}^{(\ell)}_i  + {M}^{(\ell)}_j =  M^*_i + (1 - \min_{j'} M^*_{j'})^+\ge 1$ since by definition \begin{equation*}M^*_i \ge \min_{j':v_{j'}\in {\cal V}_\ell, (\{i,j\},\ell)\in E} M^*_{j'}.\end{equation*} 
 
 \item{$i\in  {\cal V}_\ell\backslash \bigcup_{k\neq \ell}{\cal F}_{k,\ell}$ and $j\in {\cal F}_{k,\ell}$.
 We can write  ${M}^{(\ell)}_i  + {M}^{(\ell)}_j =  M^*_i + M_j^* - (1 - \min_{j'}M^*_{j'})^+$. If $1 - \min_{j'}M^*_{j'} < 0$, we trivially have $M^{(\ell)}_i  + {M}^{(\ell)}_j \ge 1$. Suppose $1 - \min_{j'}M^*_{j'}\ge 0$. For any $j'\in {\cal V}_k$ for which $(\{j,j'\},k)\in E$ we have}
\begin{eqnarray*}
 &&\hspace{-0.7cm}F(M^*_i + M_j^* + M^*_{j'}-1) \\
 &\ge&H(h_i^*) + H(h_j^*) + H(h_{j'}^*)-F\\
 &\ge&H(h_i^*,h_j^*,h_{j'}^*) -F\\
&\stackrel{(\#)}{\ge}& 2F +  H(h_i^*,h_j^*,h_{j'}^*|A_k,A_\ell)-F\ge F
\end{eqnarray*}
where $(\#)$ follows from the fact that $(h_1^*,\dots,h_K^*)$ is a valid assignment and therefore, in graph $G$ the triple $(h_i^*,h_j^*,h_{j'}^*)$ must be able to reproduce the files $A_k$ and $A_\ell$. Thus, ${M}^{(\ell)}_i  + {M}^{(\ell)}_j = M^*_i +M^*_j+\min_{j'} M^*_{j'} - 1\ge 1$.

\item{$i,j\in {\cal F}_{k,\ell}$ for some $k\neq \ell$. We can write  ${M}^{(\ell)}_i  + {M}^{(\ell)}_j =  M^*_i - (1- \min_{i'}M^*_{i'})^++ M_j^* -(1- \min_{j'}M^*_{j'})^+$. Again, if $ 1- \min_{i'}M^*_{i'}<0$ or $1- \min_{j'}M^*_{j'}<0$, the proof is simple. Suppose both these expressions are non-negative. For any $i',j'\in {\cal F}_{\ell,k}$ for which $(\{i,i'\},k),(\{j,j'\},k)\in E$ we have}
\begin{eqnarray*}
 &&\hspace{-0.68cm}F(M^*_i +M^*_{i'}+ M_j^* + M^*_{j'}-  2) \\
 &\ge&H(h_i^*,h_{i'}^*) + H(h_j^*,h_{j'}^*)-2F\\
  &\ge&2F + H(h_i^*,h_{i'}^*|A_k) + H(h_j^*,h_{j'}^*|A_k)-2F\\
 &\ge&H(h_i^*,h_j^*,h_{i'}^*,h_{j'}^*|A_k) \\
  &\ge&H(h_i^*,h_j^*|A_k)\ge F+ H(h_i^*,h_j^*|A_k,A_\ell)\ge F.
\end{eqnarray*}
And therefore, ${M}^{(\ell)}_i  + {M}^{(\ell)}_j \ge 1$. Note that we might have $i' = j'$ but this does not affect  the analysis above.
\item{$i\in{\cal F}_{k,\ell}$ and $j\in {\cal F}_{k',\ell}$ for $k\neq k'$}. The analysis is very similar to the previous case. 
\end{itemize}
Let $(M_1,\dots,M_K) = (\sum_{\ell =1}^L M^{(\ell)}_1,\dots,\sum_{\ell =1}^L M^{(\ell)}_L)$ and let $M = \sum_{k=1}^K M_k$. Since $(M_1,\dots,M_K)$ is a superposition of $L$ different (not necessarily optimal) solutions for $G_1,\dots,G_L$, clearly we have that $M \ge \hat{M}$. But we have $M = M^*$, because:
\begin{eqnarray*}
M&=&\hspace{-0.3cm} \sum_{k:v_k\in{\cal V} \backslash \bigcup_{i\neq j}{\cal F}_{i,j}}\sum_\ell M_k^{(\ell)}+\sum_{k: v_k\in {\cal F}_{i,j}, i\neq j}\sum_{\ell\in\{i,j\}}M_k^{(\ell)}\\
&=&\hspace{-0.3cm} \sum_{k:v_k\in{\cal V} \backslash \bigcup_{i\neq j}{\cal F}_{i,j}}M_k^*+\hspace{-0.3cm}\sum_{k: v_k\in {\cal F}_{i,j}, i\neq j} (1- \min_{i':v_{i'}\in {\cal V}_i\cdots} M_{i'}^*)^+ \\&+&\sum_{k: v_k\in {\cal F}_{i,j}, i\neq j}  M_k^*-(1-\min_{j':v_{j'}\in {\cal V}_i\cdots}M_{j'}^*)^+\\
&=&\sum_{k:v_k\in{\cal V} \backslash \bigcup_{i\neq j}{\cal F}_{i,j}}M_k^* + \sum_{k: v_k\in {\cal F}_{i,j}, i\neq j }M_k^* = M^*.
\end{eqnarray*}
Therefore, we established that $\hat{M}\le M^*$. Trivially, we also have that $M^*\le \hat{M}$ which proves $\hat{M} = M^*$.
\end{IEEEproof}
At a first glance, it is tempting to conjecture that the constraint $|{\cal N}_i| =1$ imposed by Theorem \ref{thm:onetoone} is not of fundamental importance and can be relaxed. Nevertheless, this is only partially true. As we will see shortly, even when $L = 2$, $|{\cal N}_1| =1$ and $|{\cal N}_2 |>1$, Algorithm \ref{Algorithm1} may fail to return an optimal solution. On the bright side, we can still make the following claim.
\begin{theorem}
Suppose a graph $G = ({\cal V},E)$ is smoothly colored with respect to a partitioning ${\cal N}_1$ and ${\cal N}_2$ where $|{\cal N}_1|=1$. Suppose further that $F_{2,j}=\varnothing$ for $j=1,2$. Let $(\hat{M}_1,\dots,\hat{M}_K) = \mbox{SUP}(G,{\cal N}_1,{\cal N}_2)$ and $({M}^*_1,\dots,{M}^*_K) = \mbox{solve}(G)$ and, $\hat{M} = \sum_{i=1}^K \hat{M}_i$ and $M^* = \sum_{i=1}^K {M}^*_i$. We have $\hat{M} = M^*$.
\label{thm:onesided}
\end{theorem}
\begin{IEEEproof}
Suppose $(h_1^*,\dots,h_K^*)$ is a valid assignment for $G$ which satisfies $H(h_\ell^*) \le M_\ell^*$. Without loss of generality, assume ${\cal N}_1 = \{1\}$ and ${\cal N}_2 = \{2,\dots,N\}$. Consider the following assignment for $G_1$. For all $v_i\in {\cal V}_1$ we set $M^{(1)}_i = M^*_i$ and for all $v_i\in {\cal F}_{1,2}$ we set $FM^{(1)}_i = I(h_i^*;A_1)$. We set $M^{(1)}_i = 0$ for all the other vertices. As for $G_2$, we propose the following assignment: for all $v_i\in {\cal V}_2$ let $FM^{(2)}_i = H(h_u^*|A_1=\bar{a})$ where $\bar{a}\in \mathbb{F}_q^n$ is the solution to
\begin{eqnarray*}
\bar{a} = \argmin_{a\in \mathbb{F}_q^n} \sum_{u\in {\cal V}_2} H(h^*_u|A_1 = a)
\end{eqnarray*}
For all the remaining vertices we set $M^{(2)}_i= 0$. Note that without loss of generality, one can assume that $\bar{a}$ is a string of zeros. (If not, one can easily modify the functions $h_u^*$ such that this property is held. This can be done without changing the required memory, and the recoverability of the files.)\\
Similar to the previous proof, we will show that these two are valid assignments. Firstly, $(M^{(2)}_1,\dots,M^{(2)}_K)$  is a valid assignment for $G_2$ because for all nodes in ${\cal V}_2$ we can store $h^{(2)}_u = h_u^*({\bf 0},A_2,\dots,A_N)$. We have $H(h_u^*({\bf 0},A_2,\dots,A_N)) = H(h^*_u|A_1 ={\bf 0})$. For all $\{u,v\}\in {\cal V}_2$ where $(\{u,v\},c)\in E$ for some $c\neq 1$, since:
\begin{equation*}
H(A_c\Big|h_u^*(A_1,A_2,\dots,A_N),h_v^*(A_1,A_2,\dots,A_N))=0
\end{equation*}
we must have that \begin{eqnarray*}H(A_c\Big|h_u^*(a,A_2,\dots,A_N),h_v^*(a,A_2,\dots,A_N),A_1=a) = 0\end{eqnarray*} for all $a\in\mathbb{F}_q^n$ including $a = {\bf 0}$. Therefore, $H(A_c\Big|h_u^*({\bf 0},A_2,\dots,A_N),h_v^*({\bf 0},A_2,\dots,A_N),A_1 = {\bf 0}) = 0$ and thus $H(A_c\Big|h_u^*({\bf 0},A_2,\dots,A_N),h_v^*({\bf 0},A_2,\dots,A_N)) = 0$ (because $A_1$ is independent of $(A_2,\dots,A_N)$).

Secondly, $(M^{(1)}_1,\dots,M^{(1)}_K)$  is a valid assignment for $G_1$ because if $u\in {\cal V}_1$ and $v\in {\cal V}_2$ and $(\{u,v\},1)\in E$ then
\begin{eqnarray*}
F(M^{(1)}_u+M^{(1)}_v) &=& FM^*_{u} + I(h_v^*;A_1)\\
&=& H(h_u^*)+H(h_v^*) -  H(h_v^*|A_1)\\
&\ge& H(h_u^*,h_v^*) - H(h_v^*,h_u^*|A_1)\\
&=& H(A_1) = F.
\end{eqnarray*}
The rest of the proof is simple. Since
\begin{eqnarray*}
\sum_{u\in {\cal V}_2} H(h_u^*|A_1 ={\bf 0}) \le \sum_{u\in {\cal V}_2} H(h^*_u|A_1)
\end{eqnarray*}
it follows that 
\begin{eqnarray*}
FM&=&F(\sum_{u\in {\cal V}} (M_u^{(1)} + M_u^{(2)})) \\
&\le&F\sum_{u\in {\cal V}_1} M^*_u + \sum_{u\in {\cal V}_2} I(h^*_u;A_1) + H(h^*_u|A_1)\\
&=&F\sum_{u\in {\cal V}_1} M^*_u + \sum_{u\in {\cal V}_2} H(h_u^*)=F\sum_{u\in {\cal V}} M^*_u = FM^*.
\end{eqnarray*}
Naturally, $\hat{M}\le M$ and therefore, $\hat{M}  = M^*$.
\end{IEEEproof}
The condition $F_{2,j}=\varnothing, j=1,2$ imposed by Theorem \ref{thm:onesided} is clearly stronger than the constraint ${\cal F}_{1,j}\cap {\cal F}_{2,j} = \varnothing, j=1,2$ from Theorem \ref{thm:onetoone}. It tells us that the cross-edges must be all colored similarly to the monochromatic cluster. 

\begin{figure}
\includegraphics[scale=0.5]{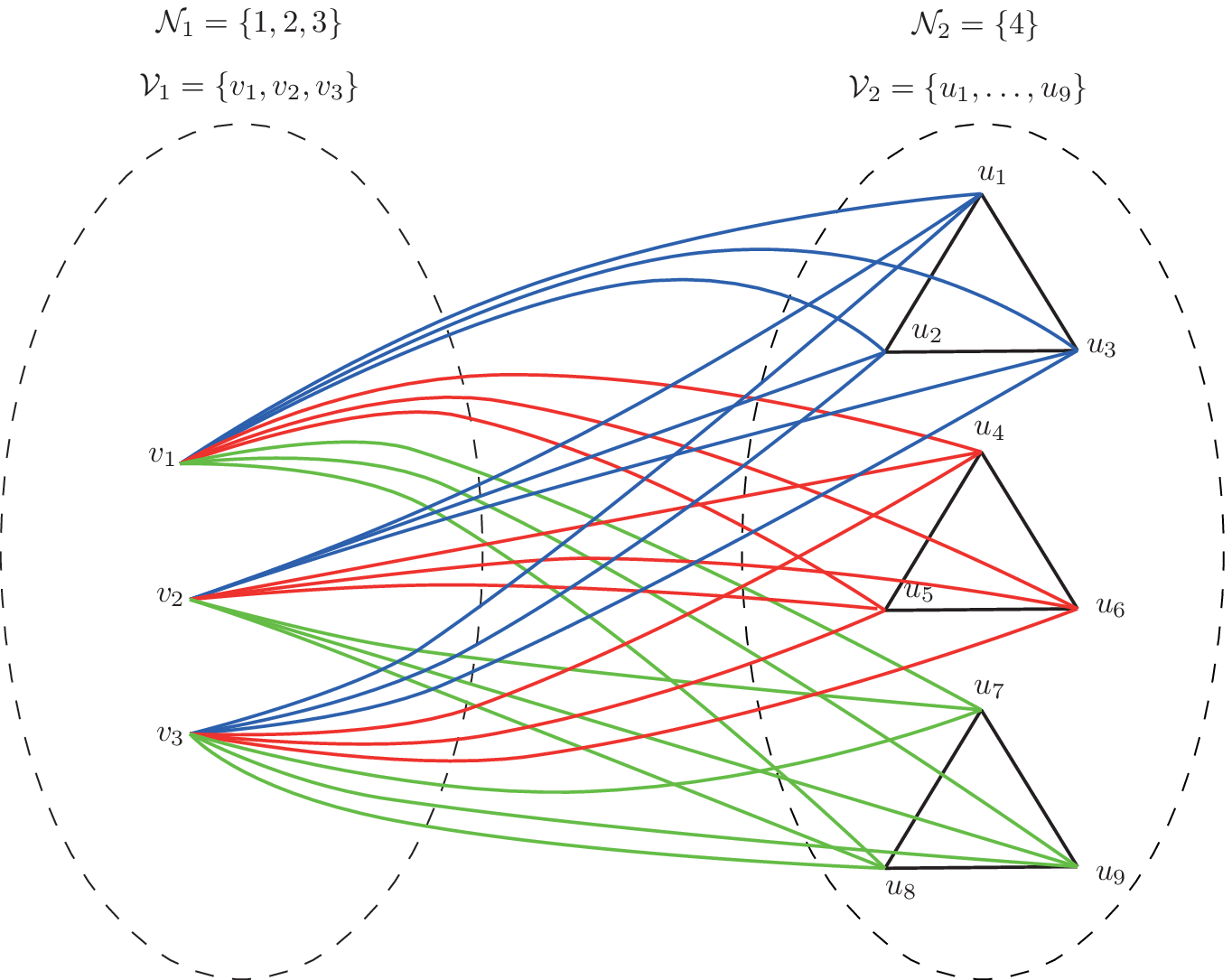}
\caption{If $|{\cal N}_1|>1$ and ${\cal F}_{1,j}\neq \varnothing$ then Algorithm \ref{Algorithm1} is suboptimal in general.}
\label{fig:counterexample}
\end{figure}

If not, Algorithm \ref{Algorithm1} may be strictly sub-optimal. An example is depicted in Figure \ref{fig:counterexample}.
This is a complete bipartite graph superimposed with the edges $(\{u_{i+3\ell},u_{j+3\ell}\},4)$ for all $\{i,j\}\subset \{1,2,3\}$ and $\ell \in \{0,1,2\}$. All the black edges are interested in $A_4$, while the blue, red and green edges are interested in $A_1$, $A_2$ and $A_3$ respectively. We show that applying Algorithm 1 on this network with partitioning ${\cal N}_1$ and ${\cal N}_2$ provides a strictly suboptimal solution. Suppose $G_1$ and $G_2$ are the two subgraphs obtained from Algorithm $1$. By applying Theorem \ref{thm:onesided} twice on $G_1$ one can find its optimal solution: $h_{v_i} = \{A_1,A_2,A_3\}$ for $i\in \{1,2,3\}$ (and all the other node store nothing). Then the solution to $G_2$ can be easily verified as $h_{u_1} = h_{u_4} = h_{u_7} = A_{4}^{(1)}$ and $h_{u_2} = h_{u_5} = h_{u_8} =A_{4}^{(2)}$ and $h_{u_3} = h_{u_6} = h_{u_9} = A_4^{(1)}+A_4^{(2)}$ where $A_4$ is assumed to have $2$ symbols $A_4^{(1)}$ and $A_4^{(2)}$ and the summations are modulo $q$. Therefore, algorithm $1$ provided a solution with $\hat{M} = 13.5$. On the other hand, we can do strictly better via the following strategy: $h_{v_i} = A_4$ for $i\in\{1,2,3\}$ and $h_{u_1} = A_1 $, $h_{u_2} =A_1 + A_4$, $h_{u_3} = A_1 + 2A_4$, $h_{u_4} = A_2 $, $h_{u_5} =A_2 + A_4$, $h_{u_6} = A_2 + 2A_4$, $h_{u_7} =A_3 $, $h_{u_8} =A_3 + A_4$, $h_{u_9} = A_3 + 2A_4$. which results in $M = 12$.
\section{Hyper-graphs with One File}
\label{sec:hyper}
In this section we briefly look at the GDSP defined over an arbitrary hyper-graph but only in the presence of one file. Suppose $G=({\cal V},{E})$ where $E$ is an arbitrary subset of the power set of ${\cal V}$. We define the concepts of valid memory allocation and normalized sum achievable rate similarly to chapter \ref{sec:simple}. We only replace \eqref{eqn:condition2} by 
\begin{eqnarray*}
H(A|h_{s_1,F},\dots,h_{s_\ell,F}) = 0 \mbox{ for all } S=  \{s_1,\dots,s_\ell\}\in E.
\end{eqnarray*}
We have the following simple proposition which directly follows from the analogy that we established between GDSP and network information flow in Section \ref{sec:nif} and the min-cut max-flow theorem \cite{ahlswede2000network}.
\begin{proposition}
\label{prop:onlysizematters}
Suppose we have a hyper-graph $G=({\cal V},E)$ with $K$ vertices. Let $M^*$ be the minimum normalized sum achievable rate for $G$. Then $M^*$ is the solution to the following LP.
\begin{eqnarray*}
M^* &=& \min_{M_1,\dots,M_K} \sum_{u=1}^K M_u \;\; s.t.\\
M_u&\ge& 0,\;\;\; \forall u\in \{1,\dots,K\},\\
\sum_{u\in S}M_u&\ge& 1,\;\;\; \forall S\in E.
\end{eqnarray*}
\end{proposition}

\section{Conclusion and Future Work}
In a future work we plan to investigate the trade-off between storage and repair bandwidth for the model studied in Section \ref{sec:hyper}. Another interesting problem will be to find a (non-trivial) generalization of the concept of smoothly colored graphs that applies to arbitrary hyper-graphs, and permits an exact solution via Algorithm \ref{Algorithm1}. 

\section*{Acknowledgement}
        \thanks{This work was supported in part by the Swiss National Science Foundation under Grant 169294.}

\bibliographystyle{IEEEtran}
\bibliography{IEEEfull,graph2}

\end{document}